\documentclass[preprint,3p,times,onecolumn]{elsarticle}
\usepackage{amsmath,amssymb,multirow,graphicx,pennames,fixltx2e}

\journal{Elsevier}

\usepackage{hyperref}

\def\d{\mathrm{d}}

\begin{document}

\begin{frontmatter}

\title{New method of particle identification with tracker detectors}

\author{Ferenc Sikl\'er}
\ead{sikler@rmki.kfki.hu}
\address{KFKI Research Institute for Particle and Nuclear Physics,
         Budapest, Hungary}

\begin{abstract}

Tracker detectors can be used to identify charged particles based on their
global $\chi$ value obtained during track fitting with the Kalman filter. This
approach builds upon the knowledge of detector material and local position
resolution, using the known physics of multiple scattering and energy loss. The
proposed method is independent of the traditional way of identification using
deposited energy. The performance for present LHC experiments is demonstrated.

\end{abstract}


\end{frontmatter}

\section{Introduction}

The momentum of a charged particle can be measured by examining the small angle
scatters of the trajectory during propagation through the detector medium or
tracker layers. For a recent application see Ref.~\cite{Grossheim:2004fp}
where the root mean square of the scattering angle distribution is computed for
each track and compared to the theoretical estimate which is proportional to
$1/\beta p$. By assuming particle type, or at high momentum ($\beta \approx
1$), $p$ can be estimated. This classical method underestimates momentum since
the particle loses energy and its momentum decreases.

The Kalman filter is widely used in present particle physics experiments for
charged track and vertex fitting and provides a coherent framework to handle
known physical effects and measurement uncertainties \cite{Fruhwirth:1987fm}.
It is equivalent to a global linear least-squares fit which takes into account
all correlations coming from process noise. It is the optimum solution since it
minimizes the mean square estimation error.
Recent studies show that this technique can be successfully used to improve
momentum resolution of particles, even in experiments without magnetic field
\cite{Ankowski:2006ts}. It is possible via the effects of multiple scattering.
If the detector is in magnetic field, the momentum of charged particles can be
obtained from the bending of the trajectory. Hence track fitting may provide
additional information that could constrain the velocity of the particle, thus
contributing to particle separation or identification.

This article is organized as follows: Sec.~\ref{sec:merit} introduces the merit
function of a track fit $\chi$ and discusses its characteristics.
Sec.~\ref{sec:physics} deals with physical effects during track propagation,
while in Sec.~\ref{sec:properties} the basic scaling properties of $\chi$ are
given. In Sec.~\ref{sec:simulation} the details of the Monte Carlo simulation
and the obtained performance are shown. This work ends with conclusions and it
is supplemented by two Appendices.

\section{The merit function of the fitted track}

\label{sec:merit}

There are various merit functions that can characterize the goodness of a track
fit: sum of the squared and properly normalized predicted ($P$), filtered ($F$)
or smoothed ($S$) residuals. It can be easily shown that for each hit $\chi^2_P
= \chi^2_F$.
The filtered residuals are uncorrelated and in the Gaussian case independent.
Hence $\sum\chi^2_F$ is chi-square distributed with $r = \left[\sum_k
\mathrm{dim}(m_k)\right] - n_p$ degrees of freedom, where
$\mathrm{dim}(m_k)$ is the dimension of the $k$th hit on track and $n_p$ is the
number of track parameters.

Tests with smoothed residuals (e.g. for outlier
removal) appear to be more
powerful \cite{Fruhwirth:1987fm}, but the correlations of these residuals
between the states have to be taken into account. Their global covariance
matrix ${\cal R}_{kl}$ between smoothed states $k$ and $l$ can be calculated
\cite{Hulsbergen:2008yv} with the recursion 
\begin{gather*}
 C^n_{k-1,l} = A_{k-1} C^n_{k,l}, \qquad k \le l \\
\intertext{and}
 {\cal R}_{kl} = V_k \delta_{kl} - H_k C^n_{k,l} H_l^T
\end{gather*}

\noindent where $C$ is the smoothed covariance matrix, $A$ is the gain matrix,
$V$ is the covariance of measurement noise, $H$ is the measurement projection
matrix.
Here we follow the notations of Refs.~\cite{Fruhwirth:1987fm,Hulsbergen:2008yv}.
The vector of smoothed
residuals is described by a multivariate Gaussian distribution with the global
covariance ${\cal R}$ obtained above. Since Kalman filtering consists of a
series of linear transformations, the smoothed residuals can be obtained from
the predicted ones by a linear transformation $r_S = B r_P$. Note that no
translation is allowed, since the average of both residuals is zero. The global
covariance matrix of predicted residuals is ${\cal R}_P$, the
covariance for $r_S$ is ${\cal R}_S = B {\cal R}_P B^T$. Thus, the expression
for the corrected sum of smoothed values is
%
\begin{equation}
 \left(\sum\chi_S^2\right)' = r_S^T {\cal R}_S^{-1} r_S = 
  (B_P r_P)^T (B_P {\cal R}_P B_P^T)^{-1} (B_P r_P) 
  = (r_P^T B_P^T) ({B_P^T}^{-1} {\cal R}_P^{-1} B_P^{-1}) (B_P r_P)
  = r_P^T {\cal R}_P^{-1} r_P = \sum\chi_P^2.
\end{equation}
 
\noindent It is clear that the correlations are transformed out and we get back
simply the predicted or filtered values: $\sum\chi^2_P = \sum\chi^2_F =
(\sum\chi^{2}_S)'$.  Hence the most straightforward quantity to calculate is the
sum $\chi^2 \equiv \sum\chi^2_P$ using predicted residuals which will be used in the remaining
part of this study.

During track propagation the mass of the tracked particle has to be assumed. In
collider experiments it is often set to the mass of the most abundantly
produced particle, the pion, or that of the muon. The obtained merit function
with mass assumption $m_0$ is
\begin{equation*}
 \chi^2(m_0) = \sum_k r_k^T R_k^{-1} r_k
\end{equation*}

\noindent where the index $k$ runs for all the measurements and $R_k$ is the
local covariance matrix for the $k$th measurement. If the largest contributions
to $R_k$ are independent in $r\phi$ and $z$ directions, $\chi^2$ can written as
%
\begin{equation}
  \chi^2(m_0) \approx
    \sum_i \left(\frac{x_i - \mu_i(m_0)}{\sigma_i(m_0)}\right)^2 =
    \sum_i \left(\frac{\sigma_i(m)}{\sigma_i(m_0)}\right)^2
            \left(\frac{x_i - \mu_i(m_0)}{\sigma_i(m)}\right)^2
  = \sum_i a_i z_i \label{eq:the_sum}
\end{equation}

\noindent where $i$ runs for all split measurements and $\sigma_i$ are the
corresponding standard deviations. The resulted sum is a
linear combination of non-centrally chi-square distributed independent random
variables $z_i$ with weights $a_i$. The distribution functions are $f_X(z_i; 1,
\lambda_i)$ where
\begin{equation*}
 a_i       = \left(\frac{\sigma_i(m)}{\sigma_i(m_0)}\right)^2, \qquad
 \lambda_i = \left(\frac{\mu_i(m) - \mu_i(m_0)}{\sigma_i(m)}\right)^2.
\end{equation*}

The sum in Eq.~\eqref{eq:the_sum} can approximated by a single rescaled
non-central chi-squared distribution $1/\alpha^2 f_X(x/\alpha^2; r, \lambda^2)$
such that
\begin{gather*}
 \alpha^2  = \frac{ \sum_i a_i^2 }{\sum_i a_i  }, \quad
 r         = \frac{(\sum_i a_i)^2}{\sum_i a_i^2} - n_p, \quad
 \lambda^2 = \sum_i \lambda_i
\end{gather*}

\noindent where $n_p$ is the number of track parameters.
For details see Appendix~\ref{sec:noncentral_approx}. If $m = m_0$, we get
$a_i = 1$, $\alpha = 1$, $\lambda = 0$, and the distribution is a chi-squared
one.
If the ratio of expected
variances $a_i$ are similar for all $i$, we get 
\begin{equation}
 \alpha^2
  \approx \Biggl\langle
   \left(\frac{\sigma_i(m)}{\sigma_i(m_0)}\right)^2 \Biggr\rangle
 \label{eq:alpha2}
\end{equation}

\noindent and $r$ is the number of split measurements decreased by the number
of track parameters.

At the same time the use of the variable $\chi \equiv \sqrt{\chi^2}$ appears to be
more practical. It is described by a scaled non-central chi-distribution
$1/\alpha f(\chi/\alpha;r,\lambda)$
and well approximated by a Gaussian with parameters
\begin{align}
 \mu_\chi    &= \alpha \sqrt{r - \frac{1}{2} + \lambda^2}, &
 \sigma_\chi &= \alpha \sqrt{\frac{1}{2}}.
 \label{eq:chi_gauss}
\end{align}

\noindent For detailed derivation see Appendix~\ref{sec:chi}.

The value of $\chi$ can be calculated for each track during the track fit with
Kalman filter. For different type of particles it will have different
distribution function, because the parameters $\mu_\chi$ and $\sigma_\chi$ (via
$\alpha$ and $r$) depend on the ratio of expected hit deviations
$\sigma_i(m)/\sigma_i(m_0)$ which are mass dependent (see
Sec.~\ref{sec:physics}). This observation allows to use this
quantity in particle identification. Using the Gaussian approximation of
Eq.~\eqref{eq:chi_gauss}, the separation power $\rho_\chi$ of $\chi$ between
particles of mass $m_1$ and $m_2$ is
\begin{equation}
 \rho_\chi = \frac{2[\mu_\chi(m_1) - \mu_\chi(m_2)]}
                  {\sqrt{\sigma_\chi^2(m_1) + \sigma_\chi^2(m_2)}}.
 \label{eq:rho_gauss}
\end{equation}

\section{Physical effects}

\label{sec:physics}


When a stable charged particle propagates through material
the most important effects which alter its momentum vector are multiple
scattering ($ms$) and energy loss ($el$). In the following the expected spatial
shift $\delta$ and deviation $\sigma$ will be calculated. They are to be
compared with the resolution of the local position measurement $\sigma_{pos}$
of the tracker layers.

\begin{figure*}

 \begin{center}
  \input{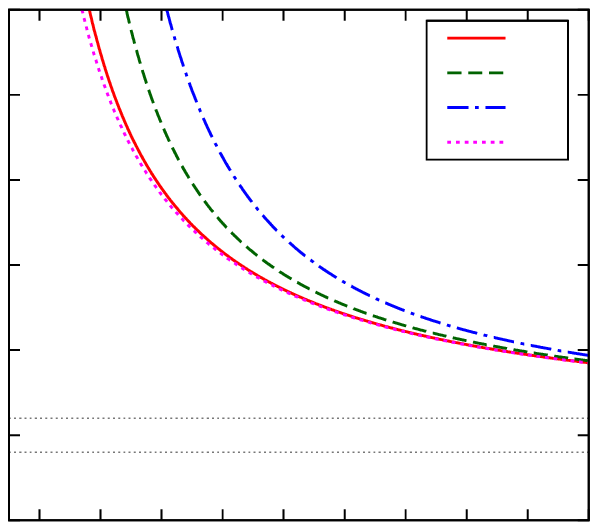}
  \input{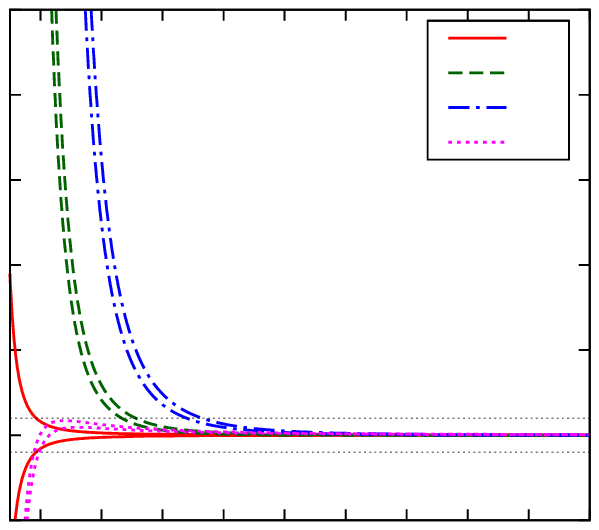}
 \end{center}

 \caption{The contributions to deviations and shifts of the predicted hit in a
$B=3.8~\mathrm{T}$ magnetic field, after crossing $x/X_0=2\%$ silicon and
further $l=5~\mathrm{cm}$ propagation before reaching the next layer, as a
function of particle momentum. Left: expected standard deviations due multiple
scattering. Right: expected shifts, compared to an average propagation
with \Pgp\ mass assumption, due to energy loss. The curves give the limits of
the lower and upper $\pm 1 \sigma$ confidence intervals for several particle
types.  For comparison lines corresponding to a local position resolution of
25~$\mu$m are drawn.}

 \label{fig:multi_eloss}

\end{figure*}

The distribution of multiple Coulomb scattering is roughly Gaussian
\cite{Amsler:2008zzb}, the standard deviation of the planar scattering angle is
\begin{equation}
 \theta_0 = \frac{13.6~\mathrm{MeV}}{\beta c p} z \sqrt{x/X_0}
  \bigl[1 + 0.038 \ln(x/X_0) \bigr] \\
 \label{eq:theta_0}
\end{equation}

\noindent where $p$, $\beta c$, and $z$ are the momentum, velocity, and charge
of the particle in electron charge units, and $x/X_0$ is the thickness of the
scattering material in radiation lengths. While the expected shift is
$\delta_{ms} = 0$, the average deviation on the next tracker plane after a
flight path $l$, in case of normal incidence, is
\begin{equation}
 \sigma_{ms} \approx l \; \theta_0.
 \label{eq:sigma_ms}
\end{equation}

Momentum and energy is lost during traversal of sensitive detector layers and
support structures. To a good approximation the most probable energy loss
$\Delta_p$, and the full width of the
energy loss distribution at half maximum $\Gamma_\Delta$ \cite{Bichsel:1988if}
are
\begin{align}
 \label{eq:mp_energyLoss}
 \Delta_p &= \xi \left[\ln\frac{2 mc^2 \beta^2 \gamma^2 \xi}{I^2}
             + 0.2000 - \beta^2 - \delta \right] \\
 \label{eq:gamma_energyLoss}
 \Gamma_\Delta &= 4.018 \xi
\end{align}
\noindent where
\begin{align*}
  \xi = \frac{K}{2} z^2 \frac{Z}{A} \rho \frac{x}{\beta^2}
\end{align*}

\noindent is the Landau parameter; $K = 4\pi N_A r_e^2 m_e c^2$; $Z$, $A$ and
$\rho$ are the mass number, atomic number and the density of the material,
respectively \cite{Amsler:2008zzb}. Since this study deals with momenta below
2~GeV/$c$, the density correction $\delta$ was neglected.

In most cases tracker detectors are placed in magnetic field ($B$). Given the
radius of the trajectory $r$ and the length of the arc $l$, the central angle
is $\varphi = l/r$. If the radius is changed by $\delta r$, the angle changes
by $\delta\varphi = -l/r^2 \; \delta r$ and the position shift of the
trajectory after $l$ path is
\begin{equation*}
 \delta_{el} \approx l \; \delta\varphi/2 = - l^2/2 \; \delta r/r^2
\end{equation*}

At the same time $p = 0.3 B r$, $E \d E = p \d p$. Hence
\begin{equation*}
 \delta_{el} \approx
   - \frac{0.3 B l^2}{2} \frac{\langle\Delta\rangle}{\beta p^2}.
\end{equation*}

Similarly, the expected deviation is
\begin{equation*}
 \sigma_{el} \approx \frac{0.3 B l^2}{2} \frac{\sigma_\Delta}{\beta p^2}.
\end{equation*}

The contributions to deviations and shifts of the predicted hit in a
$B=3.8~\mathrm{T}$ magnetic field, after crossing $x/X_0=2\%$ silicon and
further $l=5~\mathrm{cm}$ propagation before reaching the next layer, are shown
in Fig.~\ref{fig:multi_eloss}. Standard deviations are dominated by multiple
scattering, although at very low momentum the energy loss, at very high
momentum the local position measurement also plays a role. Shifts from energy
loss are only relevant at very low momentum, but they are still very small
compared to standard deviations.

\section{Properties of $\chi$}

\label{sec:properties}

It is important to study the sensitivity of the measured $\chi$ distribution at
a given total momentum $p$. The parameters which govern the distribution
(Eq.~\eqref{eq:chi_gauss}) are the rescaler $\alpha$, the average shift
$\lambda$ and the number of degrees of freedom $r$. In this section we estimate
them, as well as the separation power $\rho_\chi$ listed in
Eq.~\eqref{eq:rho_gauss}, based on physical effects.

Since the deviations are dominated by multiple scattering and local position
measurement, $\alpha$ in Eq.~\eqref{eq:alpha2} can approximated as
\begin{gather*}
 \alpha \approx \sqrt{\frac{\sigma_{ms}^2(m)   + \sigma_{pos}^2}
                           {\sigma_{ms}^2(m_0) + \sigma_{pos}^2}}
\intertext{which can be further simplified, if $\sigma_{pos} \ll \sigma_{ms}$, to}
  \alpha
  \approx \frac{\beta(m_0)}{\beta(m)}
   \left[1 - \frac{\zeta^2}{2}
             \left(1 - \frac{\beta^2(m)}{\beta^2(m_0)}\right) \right]
\end{gather*}

\noindent where the sensitivity is defined as $\zeta =
\sigma_{pos}/\sigma_{ms}(m_0)$, it is proportional to $1/\beta p$
(Eqs.~\eqref{eq:theta_0}--\eqref{eq:sigma_ms}).  If the local position
resolution can be neglected ($\zeta\ll~1$) we get
\begin{equation}
 \alpha \approx \frac{\beta(m_0)}{\beta(m)}.
 \label{eq:alpha_approx}
\end{equation}

Shifts come entirely from differences in energy loss, hence contributions to
$\lambda$ are only substantial at low momentum:
%
\begin{equation}
 \lambda \approx
  \sqrt{r} \;
   \frac{0.3 B l^2 \langle\Delta(m) - \Delta(m_0)\rangle}{2 \beta p^2 l \theta_0}
   \propto \frac{l \sqrt{r x}}{p}
     \left(\frac{1}{\beta^2(m)} - \frac{1}{\beta^2(m_0)}\right).
\end{equation}

\begin{figure}
 \begin{center}
  
 \begin{center}
  \input{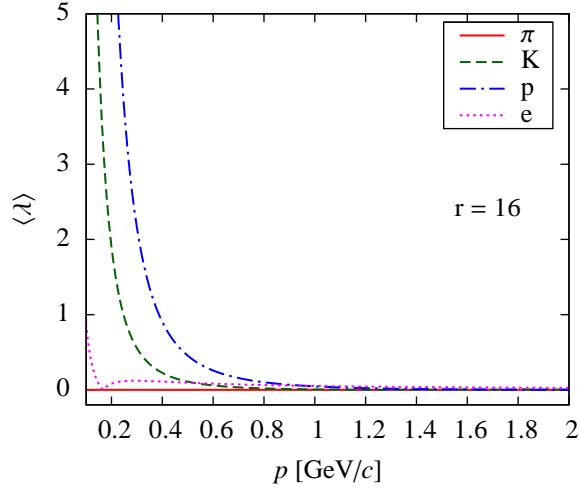}
 \end{center}
 
 \end{center}

 \caption{The average shift for several particle types, in a $B=3.8~\mathrm{T}$
magnetic field, with layer thicknesses of $x/X_0=2\%$ silicon, an average
propagation length of $l=5~\mathrm{cm}$, in case of $r = 16$ number of degrees
of freedom, as a function of the particle momentum.}

 \label{fig:lambda}

\end{figure}

\noindent The average shift $\langle\lambda\rangle$ in a $B=3.8~\mathrm{T}$
magnetic field, with layer thicknesses of $x/X_0=2\%$ silicon, an average
propagation length of $l=5~\mathrm{cm}$, in case of $r = 16$ number of degrees
of freedom, is shown in Fig.~\ref{fig:lambda}.

If $\lambda,\zeta\ll~1$, the separation power $\rho_\chi$ between particles $m$
and $m_0$ is
\begin{equation}
 \label{eq:separationPower}
  \rho_\chi \approx 2\sqrt{2r-1} \;
   \frac{1 -  \beta(m)/\beta(m_0)} {\sqrt{1 + [\beta(m)/\beta(m_0)]^2}}.
\end{equation}

\noindent Hence if the momentum is not very low and the local position
resolution is small compared to deviations from multiple scattering, neither
the rescaler $\alpha$ nor the separation power $\rho_\chi$ depends on the
details of the experimental setup, such as magnetic field, radii of tracker
layers, value of local position resolution and material thickness. In this
respect the only decisive parameter is the number of split measurements which
enters the above expressions by the number of degrees of freedom $r$.  The mean
and variance of the corresponding Gaussians are fully determined by the
momentum and mass of the particles via $\beta$.

Although at low momentum the prediction of the means is more difficult due to
the increasing $\lambda$, the variances still stay the same. The $\chi$
distribution can be easily unfolded, since the separation power is large,
allowing for a many-parameter fit.

%
%
%

\subsection{Applications}

The measured value of $\chi$ is sensitive to the proper spatial alignment of
the detector layers and to the correct estimate of the variation of the
predicted local position. If the alignment precision is sufficient, the latter
is mostly determined by the contribution from multiple scattering which is
closely proportional to $\sqrt{x/X_0}$. While $p$ and $r$ are well measured,
the amount of material in the detector can be

\begin{table*}

 \caption{Important characteristics of the inner barrel detectors of the
studied experimental setup. For details see text at the beginning of
Sec.~\ref{sec:simulation}.}


 \label{tab:experiments}

 \begin{center}
 \begin{tabular}{ccllccccccc}
  \hline
        & $B$ & \multicolumn{1}{c}{Subdetector}
              & \multicolumn{1}{c}{Radius of layers}
              & $\sigma_{r\phi}$ & $\sigma_z$
              & $x/X_0$ & $\zeta_{r\phi}$ & $\zeta_z$
              & Split \\
  & [T] & & \multicolumn{1}{c}{[cm]} & [$\mu$m] & [$\mu$m] & [\%] & & & meas. \\
  \hline
  \multirow{3}{*}{Exp A} & \multirow{3}{*}{2}
    & pixels (barrel)  & 5.0, 8.8, 12.2         &  10 & 115 & 4   & 0.1 & 1 
      & \multirow{3}{*}{50} \\
  & & strips (SCT)$^s$ & 29.9, 37.1, 44.3, 51.4 &  17 & 580 & 4   & 0.1 & 3 \\
  & & straw (TRT) & 56.3 -- 106.6 ($\le$ 36 hits)
                                                 & 130 &  -- & 0.5 & 10 & -- \\
  \hline
  \multirow{4}{*}{Exp B} & \multirow{4}{*}{0.4}
    & pixels (SPD)     &  3.9,  7.6 & 12 & 100 & 1 & 0.2 & 2 
      & \multirow{4}{*}{12} \\
  & & drifts (SDD)     & 14.9, 23.8 & 35 &  23 & 1 & 0.3 & 0.2 \\
  & & strips (SSD)$^s$ & 38.5, 43.6 & 15 & 730 & 1 & 0.1 & 7 \\
  & & [gas (TPC)        & 84.5 -- 246.6 ($\le$ 159 hits)
                       & 900 & 900 & $10^{-3}$ 
                       & \multicolumn{2}{c}{$10^3-10^4$]}  \\
  \hline
  \multirow{5}{*}{Exp C} & \multirow{5}{*}{3.8}
    & pixels (PXB)     & 4.4, 7.3, 10.2      & 15            &  15 & 3
                       & 0.2 & 0.2 & \multirow{5}{*}{20} \\
  & & strips (TIB)$^s$ & 25.5, 33.9          & 23/$\sqrt{2}$ & 230 & 4
                       & 0.1 & 0.8 \\
  & & strips (TIB)     & 41.8, 49.8          & 35            &  -- & 2
                       & 0.2 & -- \\
  & & strips (TOB)$^s$ & 60.8, 69.2          & 53/$\sqrt{2}$ & 530 & 4
                       & 0.1 & 2 \\
  & & strips (TOB)     & 78.0, 86.8, 96.5, 108.0    & 53, 35 &  -- & 2
                       & 0.2 & -- \\
  \hline
 \end{tabular}
 \end{center}
\end{table*}

\begin{itemize}

 \item {\it understood}: the unfolding of the $\chi$ distribution in a phase
space bin enables the measurement of yields of different particle species.

 \item {\it poorly known}: the unfolding of the $\chi$ distribution in a phase
space bin may provide corrections to the material thickness. They can be
extracted by fitting the $\chi$ distribution with an additional rescaler.  Note
that the measurement of yields of different particle species is still possible,
although with lower confidence.

\end{itemize}

\section{Simulation}

\label{sec:simulation}

The proposed method was verified by a Monte Carlo simulation. As examples from
LHC, the performance of simplified models for the inner detectors of the
following experiments were studied:

\begin{itemize}

 \item ATLAS (Exp~A): three layers of silicon pixels, five layers of
double-sided silicon strips, up to 36 layers of straw tubes \cite{Aad:2009wy}.

 \item ALICE (Exp~B): two layers of silicon pixels, two layers of silicon
drifts and two layers of double-sided silicon strips
\cite{Antinori:2003dg,Antonczyk:2006qi}. Due to the large $\zeta$ value of the
gas detector (TPC) its measurements were not included.

 \item CMS   (Exp~C): three layers of silicon pixels, ten layers of silicon
strips (four of them double-sided) \cite{Adolphi:2008zzk}.

\end{itemize}

Some relevant details of the experimental setups are given in
Table~\ref{tab:experiments}.  For simplicity a homogeneous longitudinal
magnetic field was used, and detector layers were assumed to be concentric
cylinders around the beam-line. Pixels, double-sided strips (superscript~$^s$),
drift layers and gas provide measurements in two dimensions ($r\phi$ and $z$),
while one-sided strips and straw tubes give only measurement in one direction
($r\phi$). $x/X_0$ values are given per layer and they are rounded to integers
where possible.  Sensitivity values $\zeta_{r\phi}$ and $\zeta_z$ are shown for
pions at $p = 1~\mathrm{GeV}/c$, normal incidence, rounded to one significant
digit. The number of split measurements are also indicated.

\begin{figure*}[!t]

 \begin{center}
  \input{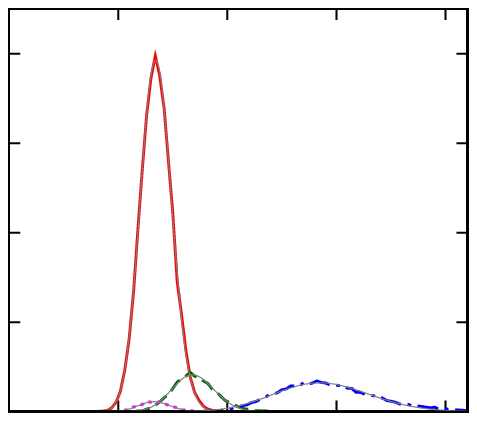}
  \input{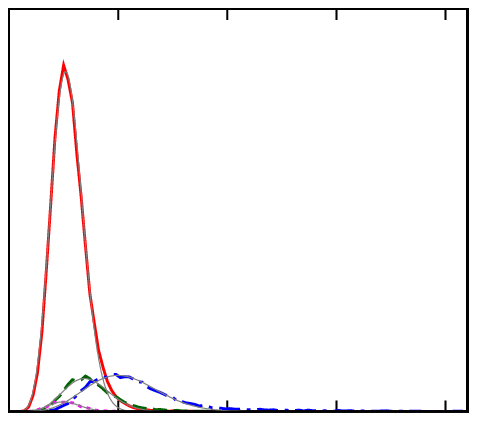}
  \input{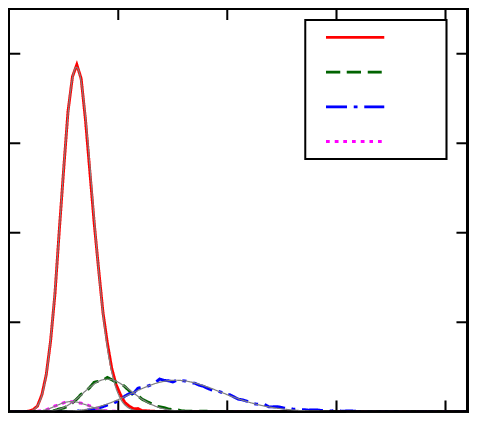}
 \end{center}

 \begin{center}
  \input{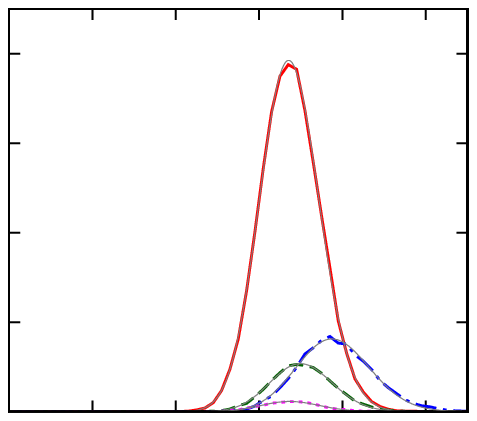}
  \input{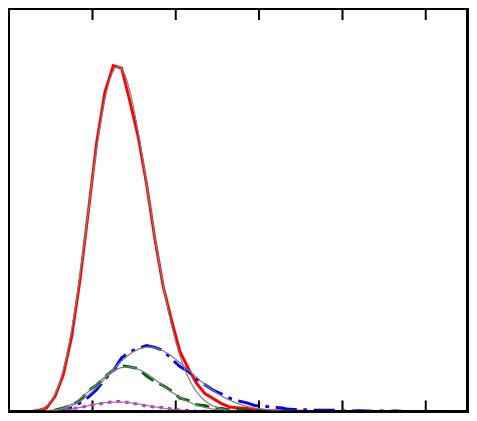}
  \input{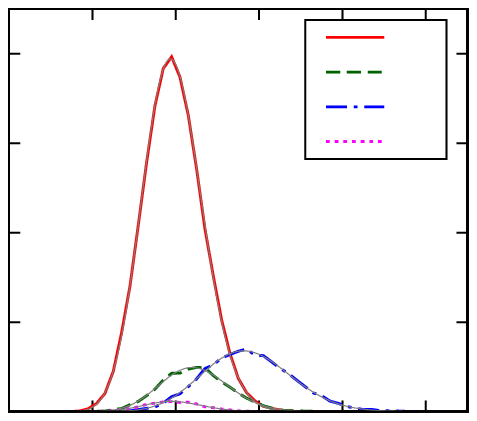}
 \end{center}

 \medskip
 \caption{Distributions of $\chi$ for several particle species. The relative
yield of particles was set to $\Pgp:\PK:\Pp:\Pe = 70:10:18:2$. Results are
shown for $\eta=0, p_T = 0.4~\mathrm{GeV}/c$ (upper row) and $\eta=0, p_T =
0.8~\mathrm{GeV}/c$ (lower row) with setups Exp A, B and C. Individual fits
with chi distributions are indicated by thin solid lines.}

 \label{fig:chi}

\end{figure*}

The initial state vector was estimated by fitting a helix to the first three
hits. (These hits are two-dimensional in all three examined experimental
setups.) The starting values of the track parameters were extracted at the
closest approach to the beam line.  The track fitting was performed by a
classical Kalman filter \cite{Fruhwirth:1987fm} with pion mass assumption. The state vector $x =
(\kappa, \theta, \psi, r\phi, z)$ is five dimensional, where
\begin{align*}
 \kappa &= q/p             & \text{(signed inverse momentum)} \\
 \theta &= \theta(\vec{p}) & \text{(local polar angle)} \\
 \psi   &= \phi({\vec p})  & \text{(local azimuthal angle)} \\
 r\phi  &= r\phi({\vec r}) & \text{(global azimuthal position)} \\
 z      &= r_L             & \text{(global longitudinal position)}.
\end{align*}

\noindent
The propagation from layer to layer was calculated analytically using a helix
model. Multiple scattering and energy loss in tracker layers was implemented
with their Gaussian approximations shown in
Eqs.~\eqref{eq:sigma_ms}--\eqref{eq:gamma_energyLoss}. The propagation matrix $F = \partial f /
\partial x$ was obtained by numerical derivation. The measurement vector $m =
(r\phi, z)$ is two dimensional, the measurement operator is 
\begin{equation*}
  H = \begin{pmatrix}
       0 & 0 & 0 & 1 & 0 \\ 0 & 0 & 0 & 0 & 1
      \end{pmatrix}.
\end{equation*}

\noindent
The covariance of the process noise $Q$ is
\begin{gather*}
 Q = (F_\kappa \otimes F_\kappa^T) \sigma_\kappa^2 +
     (F_\theta \otimes F_\theta^T) \sigma_\theta^2 +
     (F_\psi   \otimes F_\psi^T)   \sigma_\psi^2
\end{gather*}

\noindent where $\sigma_\kappa = \kappa \sigma_\Delta/\beta$, $\sigma_\theta =
\sigma_\psi = \theta_0$ and $F_a = \partial f/\partial x_a$ is a vector.  The
covariance of measurement noise $V$ is
\begin{equation*}
 V = \begin{pmatrix}
      \sigma_{r\phi}^2 & 0 \\
      0 & \sigma_{z}^2
     \end{pmatrix}
\end{equation*}

\noindent Note that multiple scattering contributes equally to the variation of
$\theta$ and $\psi$, while energy loss affects only $\kappa$.

\subsection{Results}

In order study the performance of $\chi$, charged pions, kaons, protons and
electrons with random azimuthal angle were generated and emitted normal to the
line of the colliding beams ($\eta = 0$) and run through the above outlined
reconstruction.

Distributions of $\chi$ using $10^5$ particle tracks are shown in
Fig.~\ref{fig:chi} for $p_T = 0.4$ and $0.8~\mathrm{GeV}/c$. For a realistic
particle composition the relative yields were set to $\Pgp:\PK:\Pp:\Pe =
70:10:18:2$.  At $p_T = 0.4~\mathrm{GeV}/c$, in case of Exp A, the protons are
detached, but there is a good \Pgp--\Pp\ separation for Exp B and C, as well.
For Exp A and C the \Pgp--\PK\ separation allows for yield estimation. Even at
$p_T = 0.8~\mathrm{GeV}/c$ the observed resolution is enough to extract the
protons. When fitting the histograms a sum of chi distributions was employed
(thin solid lines), but a sum of Gaussians may also be sufficient.

\label{sec:fits}

\begin{figure*}

 \begin{center}
  \input{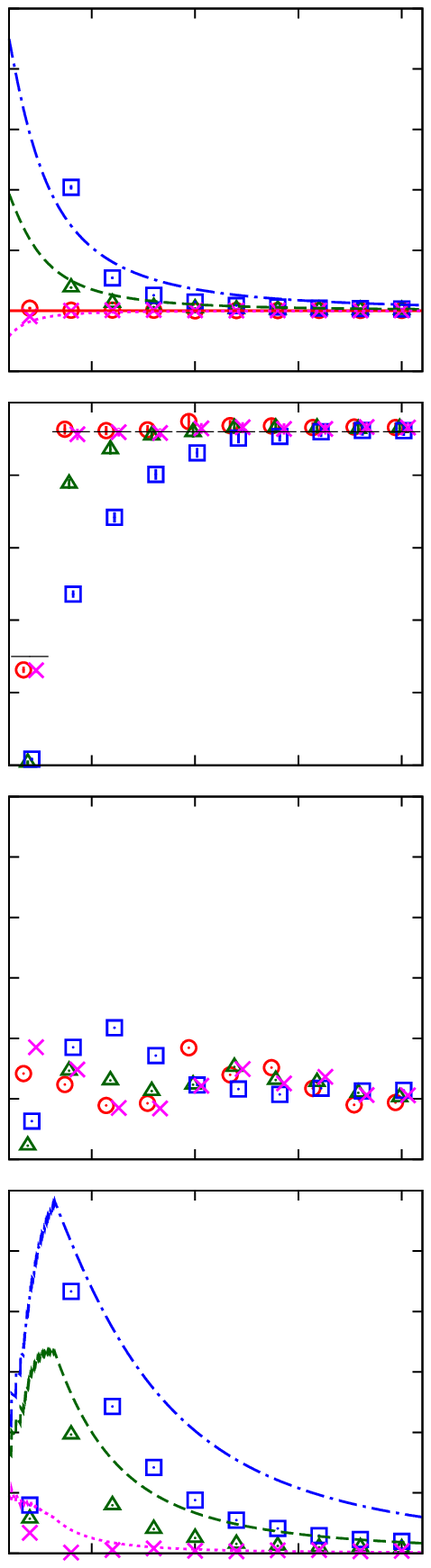}
  \input{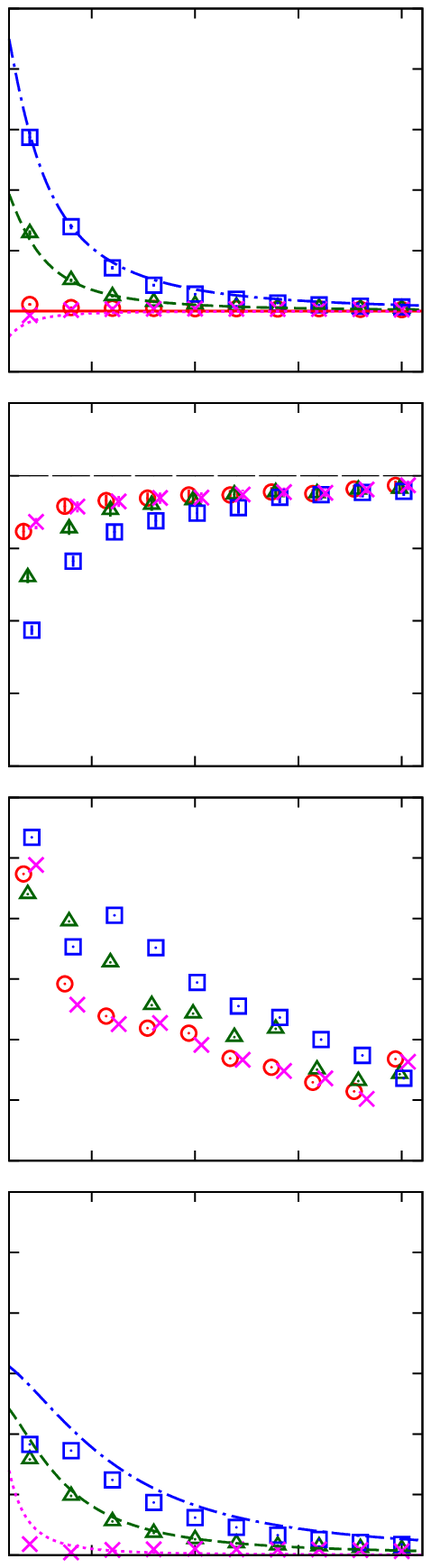}
  \input{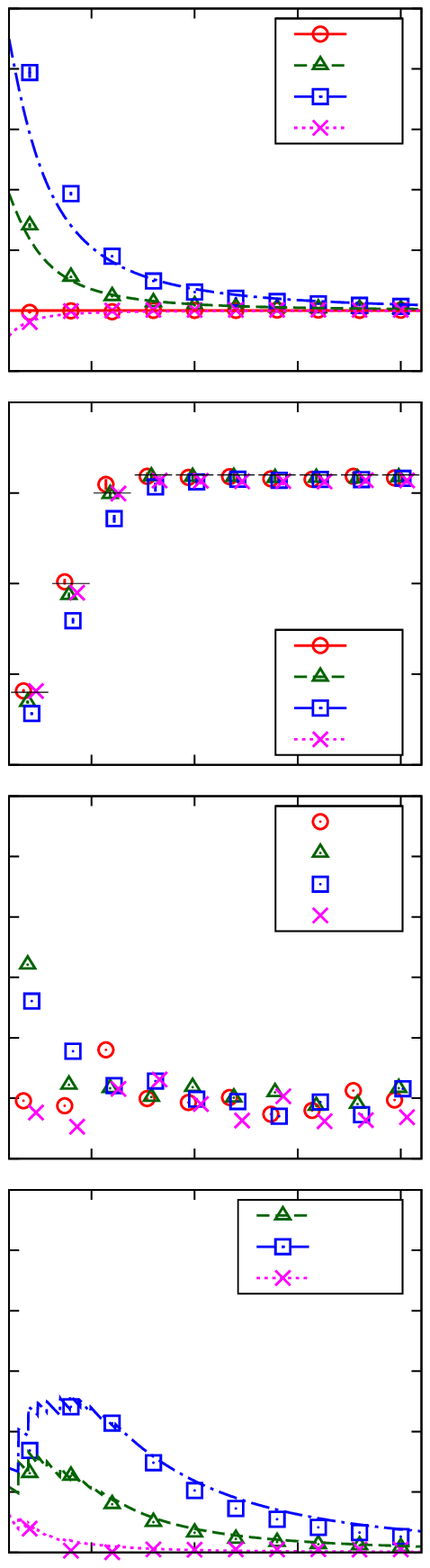}
 \end{center}

 \medskip
 \caption{Performance of $\chi$ measurement for particle identification, with
setups Exp A, B and C. For details see text in Sec.~\ref{sec:fits}.}

 \label{fig:fits}

\end{figure*}

For a complete picture charged pions, kaons, protons and electrons with
transverse momenta $p_T = 0.2$, $0.4$, $0.6$, $\dots$, $2.0~\mathrm{GeV}/c$ and $\eta =
0$ were used, amounting to $10^4$ particles per $p_T$ setting for each particle
type and experimental setup. The performance of $\chi$ as function of $p$ for
all three setups is shown in Fig.~\ref{fig:fits}. The subsequent rows give the
dependence of the measured rescaler $\alpha$, the fitted number of degrees of
freedom $r$, the merit function of the histogram fit with sum of chi
distributions $\chi_{fig}^2$ and the separation power $\rho_\chi$. This latter
was calculated by using the measured $\alpha$ and $r$ values with help of
Eqs.~\eqref{eq:chi_gauss}--\eqref{eq:rho_gauss}. The measured values are
shown by the symbols.
In case of $\alpha$ the line gives the plain $\beta(m_0)/\beta(m)$ scaling
(Eq.~\eqref{eq:alpha_approx}) that works rather well for all three setups and
for all particle types.
For $r$ the horizontal lines show the number of split measurements for a given
$p_T$, decreased by the number of track parameters $n_p$. While these
predictions are closely followed by the measured values in case of Exp C, there
are substantial deviations with the other two setups. It can be traced back to
low sensitivity measurements: large number of straw tubes with $\zeta_{r\phi} =
10$ (Exp A), and two strip layers with $\zeta_z = 7$ (Exp B).
In case of the separation power $\rho_\chi$ the lines show the approximation
based on the predicted number of degrees of freedom and the ratio
$\beta(m)/\beta(m_0)$, calculated with help of Eq.~\eqref{eq:separationPower}.
The steps are due to the changing number of crossed detector layers with
varying $p$.
The approximation works well for Exp C, but strongly overestimates the measured
value for Exp A. It is again due to the large number of low sensitivity
measurements.

\begin{figure}

 \begin{center}
  \input{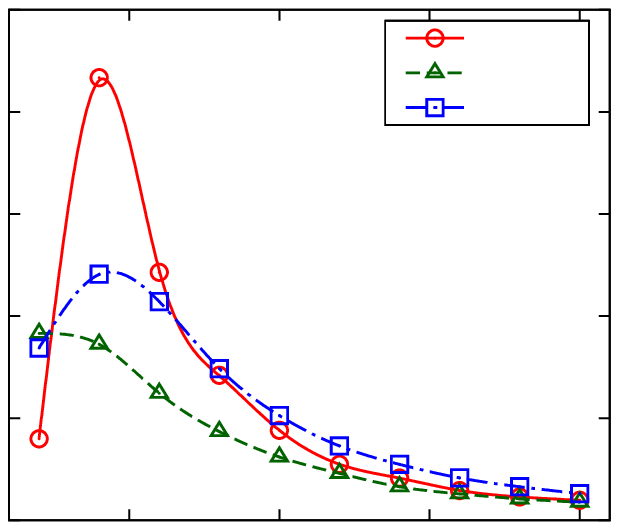}
 \end{center}

 \caption{The \Pgp--\Pp\ separation power of the $\chi$ measurement for the
experimental setups, as a function of momentum. The lines are drawn to guide
the eye.}

 \label{fig:comparison}

\end{figure}

Comparison of the \Pgp--\Pp\ separation power of the $\chi$ measurement for
several experimental setups as a function of momentum is shown in
Fig.~\ref{fig:comparison}. While Exp A clearly performs better for $p <
0.6~\mathrm{GeV}/c$, Exp C has better resolution for the more critical higher
momentum region.
With the most sensitive setups (Exp A and C) protons are $1\sigma$ apart if
$p < 1.4~\mathrm{GeV}/c$, while $2\sigma$ separation is reached if $p <
1~\mathrm{GeV}/c$. For kaons these numbers are $p < 0.9$ and
$0.5~\mathrm{GeV}/c$, respectively.

\section{Conclusions}

It was shown that tracker detectors can employed to identify charged particles
based on their global $\chi$ obtained during track fitting with the Kalman
filter. This approach builds upon the knowledge of detector material and local
position resolution, using the known physics of multiple scattering and energy
loss. The study using simplified models of present LHC experiment shows that
\Pgp--\PK\ and \Pgp--\Pp\ unfolding is possible at low momentum. The separation
is better than $1\sigma$ for $p < 0.9$ and $1.4~\mathrm{GeV}/c$, respectively.
In general, the performance of an experiment is determined by the number of good
sensitivity split measurements. It is also a strong function of particle
momentum.

If particles can be identified based on informations from other sources (e.g.
independent $\d E/\d x$ measurement) this tool can still be useful to provide
corrections to the amount of material in the detector and to check the obtained
precision of its alignment.

\section*{Acknowledgements}

The author wishes to thank to Kriszti\'an Krajcz\'ar for helpful discussions.
This work was supported by the Hungarian Scientific Research Fund and the
National Office for Research and Technology (K 48898, H07-B 74296).

\appendix

\section{Properties of some distributions}

\label{sec:chi}

In this section the definitions of some used distributions are listed along
with their calculated or approximated values for the mean $\mu$ and
variance $\sigma^2$.

\subsection{$\chi^2$ distribution}

The distribution, mean and variance are
\begin{gather*}
 P(x;r) = \frac{x^{r/2-1} e^{-x/2}}{\Gamma\left(\frac{r}{2}\right) 2^{r/2}} \\
 \mu = r,  \qquad
 \sigma^2 = 2r.
\end{gather*}

\subsection{Non-central $\chi^2$ distribution}

The distribution, mean and variance are
\begin{gather*}
 P(x;r,\lambda) = \frac{e^{-(x+\lambda)/2} x^{(r-1)/2} \sqrt\lambda}
                  {2(\lambda x)^{r/4}} \; I_{r/2-1}\left(\sqrt{\lambda x}\right) \\
 \mu = r + \lambda,  \qquad
 \sigma^2 = 2(r + 2\lambda).
\end{gather*}

\noindent where $I_n(x)$ is the modified Bessel function of the first kind.

\subsection{$\chi$ distribution}

The distribution and mean are
\begin{gather}
 P(x;r) = \frac{2^{1-n/2} x^{n-1} e^{-x^2/2}}{\Gamma\left(\frac{n}{2}\right)} \\
  \mu = \frac{\sqrt{2} \; \Gamma\left(\frac{r+1}{2}\right)}{\Gamma\left(\frac{r}{2}\right)}
   = \sqrt{r}\left[1 - \frac{1}{4r} + {\cal O}\left(\frac{1}{r^2}\right)\right]
 \label{eq:chi_mean}
   \approx \sqrt{r - \frac{1}{2}} \\
\intertext{where Ref.~\cite{Graham:1994} for $r \gg 1$ was used. The variance is}
 \label{eq:chi_sigma}
 \sigma^2 = r - \mu^2 \approx \frac{1}{2}.
\end{gather}

\subsection{Non-central $\chi$ distribution}

The distribution and mean are
\begin{gather*}
 P(x;r,\lambda) = \frac{e^{-(x^2+\lambda^2)/2} x^r \lambda}
                            {(\lambda x)^{r/2}} \; I_{r/2-1}(\lambda x) \\
 \mu = \sqrt{\frac{\pi}{2}} \; L_{1/2}^{(r/2-1)}\left(\frac{-\lambda^2}{2}\right)
\intertext{where $L_n^{(a)}(x)$ is the generalized Laguerre function.
For $r \gg 1$, with Kummer's second formula \cite{Koepf:1998}}
  \mu = \sqrt{\frac{\pi}{2}} \;
         \frac{\Gamma\left(\frac{r+1}{2}\right)}
              {\Gamma\left(\frac{r}{2}\right)\Gamma\left(\frac{3}{2}\right)} \;
         {}_1F_1\left(-\frac{1}{2}, \frac{r}{2}, \frac{-\lambda^2}{2}\right)
\end{gather*}

\noindent where ${}_1F_1(a,b,z)$ is the confluent hypergeometric function of
the first kind. With help of Eq.~\eqref{eq:chi_mean} and
Ref.~\cite{Abramowitz+Stegun}, assuming $\lambda^2 \ll r$
%
\begin{equation}
 \mu = \sqrt{r} 
  \left[1 - \frac{1}{4r} + {\cal O}\left(\frac{1}{r^2}\right)\right] 
  \left[1 + \frac{\lambda^2}{2r} + {\cal O}\left(\frac{1}{r^2}\right)\right]
\approx 
 \label{eq:noncentral_chi_mean}
  \sqrt{r} \left[1 - \frac{1 - 2\lambda^2}{4r} + {\cal
O}\left(\frac{1}{r^2}\right)\right] \approx \sqrt{r - \frac{1}{2} + \lambda^2}.
\end{equation}

For the variance
\begin{equation}
 \label{eq:noncentral_chi_sigma}
 \sigma^2 = r - \mu^2 + \lambda^2 \approx \frac{1}{2}.
\end{equation}

%

Note that with $\lambda = 0$ we get back the mean of the $\chi$ distribution
(Eqs.~\eqref{eq:chi_mean} and \eqref{eq:noncentral_chi_mean}),
while the variances are the same in the central and non-central case
(Eqs.~\eqref{eq:chi_sigma} and \eqref{eq:noncentral_chi_sigma}).


\section{Sum of non-central chi-squared distributed independent variables}
\label{sec:noncentral_approx}

The goal is to approximate the sum 
\begin{gather*}
 y = \sum_{i=1}^n a_i z_i
\end{gather*}
 
\noindent where $z_i$ are non-central chi-squared distributed independent
random variables with one degree of freedom and density function $f_X(z_i;
1,\lambda_i)$.  Although an explicit expression for the distribution of $y$
exists, it is difficult to evaluate in practice \cite{Gabler:1987}. Here this
function is approximated by a rescaled non-central chi-squared distribution
$1/\alpha f_X(x/\alpha; r, \lambda)$ by requiring that the first two moments be
the same. The means and variances are additive, thus the equations two solve
are
\begin{align*}
 \langle y \rangle &= \;\,\,  \sum_i a_i   (1 + \;\,\, \lambda_i)
                    = \;\,\,  \alpha(r + \lambda) \\
 \langle (y - \langle y \rangle)^2 \rangle
                   &= 2 \sum_i a_i^2 (1 + 2 \lambda_i)
                    = 2 \alpha^2(r + 2\lambda)
\end{align*}
 
\noindent
By assuming $\lambda_i \ll 1$ we get
\begin{gather*}
 \alpha  = \frac{ \sum_i a_i^2 }{\sum_i a_i  }, \qquad
 r       = \frac{(\sum_i a_i)^2}{\sum_i a_i^2}, \qquad
 \lambda = \sum_i \lambda_i
\end{gather*}

\noindent with relative corrections of the order ${\cal O}(\lambda^2/r^2)$.  If
the values of $a_i$ are similar some of the above expressions can be
approximated by
\begin{gather*}
 \alpha  \approx \langle a_i \rangle, \qquad
 r       \approx n.
\end{gather*}

\bibliographystyle{elsarticle-num}
\bibliography{trackerPid}

\end{document}